\documentclass[12pt]{article}
\usepackage{amssymb}
\begin{document}

\thispagestyle{empty}

\author{M.I.~Kalinin\thanks{kalinin@vniims.ru} \ and V.N.~Melnikov\thanks{melnikov@phys.msu.ru, melnikov@vniims.ru}}
\title{On the cosmological constant in quantum cosmology}

\date{}
\maketitle

\begin{abstract}
Quantization of a dust-like closed isotropic cosmological model
with a cosmological constant is realized by the method of B.
DeWitt \cite{1}. It is shown that such quantization leads to
interesting results, in particular, to a finite lifetime of the
system, and appearance of the Universe in our world as penetration
via the barrier. These purely quantum effects appear when
$\Lambda>0$.
\end{abstract}

Quantization of cosmological models becomes an important subject
on the interface of gravitation and cosmology. It was initiated by
B. DeWitt \cite{1} who performed quantization of a Friedmann
universe filled with dust. Later, closed isotropic models with
matter in the form of a conformal \cite{2} and minimally coupled
\cite{3} scalar fields were quantized. Misner's paper \cite{4} was
devoted to the problem of quantization of anisotropic cosmological
models, and Yu.N. Barabanenkov's to quantization of the Friedmann
metric matched with the Kruskal one \cite{5}.

In this paper, quantization of a dust-like closed isotropic cosmological
model with a cosmological constant is realized by a method similar to
Ref.\,\cite{1}. The system of units is chosen so that $c = \hbar = 16\pi G =
1$. The corresponding Lagrangian is
$$
    L  =  L_g + L_m
    = 12\pi ^2{\alpha}R - 12\pi ^2{\alpha}^{-1}R\
        (R_{,0})^2 -4\pi ^2{\alpha}{\Lambda}R^3+L_m,
$$
where $R$ is the radius (scale factor) of the Universe, $\Lambda$
is the cosmological constant and $-\alpha^2 = g_{00}$. $L_m$ is
the matter Lagrangian (for $N$ non-interacting particles).
Passing, by the standard recipe, to the Hamiltonian and using the
constraint equation \cite{1}, we arrive at the condition
\begin{equation}
\label{eq:486}
    {\cal H}_g+{\cal H}_m = 0,
\end{equation}
 where ${\cal H}_g = -{\Pi}^2/(48R) - 12\pi ^2R + 48\pi
^2{\Lambda}R^3$;\ ${\cal H}_m = Nm$;\ ${\Pi}$ is a momentum
conjugate to $R$. In the quantum case (\ref{eq:486}) becomes a
condition for the state vector $\Psi$
$$
    ({\cal H}_g+{\cal H}_m)\ {\Psi} = 0.
$$
All information about our system is contained in this quantum constraint
equation. Using an operator ordering such that the differential operator be
the one-dimensional Laplace-Beltrami operator, we get the following equation
in the $R$ representation:

$$
    \biggl( \frac{1}{48\pi ^2}R^{-1/4}\frac{\partial}{{\partial}R}
        R^{-1/2}\frac{\partial}{{\partial}R}R^{-1/4}-12\pi ^2R
     +4\pi ^2{\Lambda}R^3 + Nm \biggr) \Psi = 0.
$$
Here $m$ is the mass operator. The state function $\Psi$ does not depend on
$\alpha$ due to the constraint equation. Supposing that $\Psi$ is an
eigenfunction of the $m$ operator, so that we treat it here as a c-number,
and using the transformation $\chi = R^{3/2},\ \Phi = R^{-1/4}\ \Psi$,  we
come to the equation
$$
   -(3/64\pi ^2)\ \ddot{\Phi} + (12\pi ^2 \chi^{2/3}-4 \pi^2{\Lambda}
        \chi^2)\ {\Phi} = Nm{\Phi},
$$
which differs from DeWitt's equation for dust by the form of the potential.
We impose DeWitt's boundary condition $\Phi(0) = 0$, or $\Psi(0) = 0$. So, we
obtain the Schr\"odinger equation for a particle with mass $32\pi^2/3$
moving with energy $Nm$ in the one-dimensional potential
$$
    V(\chi) = \left\{   \begin{array}{cc}
            \infty, & \chi < 0, \\
     12\pi ^2\chi^{2/3}-4\pi ^2{\Lambda}\chi^2, & \chi \geq 0. \\
        \end{array} \right.
$$

There are three possible cases depending on the value of the cosmological
constant: 1) $\Lambda<0$, 2) $\Lambda = 0$, 3)$\Lambda>0$. Let us consider each
of them.

\medskip
2) The case $\Lambda = 0$ is treated in \cite{1} in the WKB approximation;
it has been obtained that the total energy is
\begin{equation}
\label{E0}
     Nm = [48\pi^2(n+3/4)]^{1/2},\quad n = 0,\ 1,\ 2,\ {\ldots}
\end{equation}
For each $Nm$ there is a maximal radius of the Universe $R_{\max}$
after which $\Psi$ exponentially decreases.

\medskip
1) $\Lambda < 0$. In this case there is also a maximal Universe
radius and, in particular, for small $Nm$, i.e., $R_{\max} \ll
(-3/\Lambda)^{1/2}$ the same expression (\ref{E0}) is obtained as
in the case $\Lambda = 0$.

For large $Nm$, i.e., $R_{\max} \gg (-3/\Lambda)^{1/2}$, as is seen from
the expression for the potential,
$$
    Nm \sim (n+3/4)\,\sqrt{-\Lambda},
$$
which is similar to the harmonic oscillator energy. The motion of
the equivalent particle will also be finite; $0 \leq R \leq
R_{\max}$.  So, for $\Lambda<0$ there are no essential differences
from the case $\Lambda = 0$.  Besides, for $\Lambda < 0$, the
results correspond to the classical case.

3) $\Lambda > 0$. In this case $V$ has the form of a potential barrier
which at large $\chi$ does not differ from a parabolic one. For
$Nm \geq 8\pi^2 \lambda^{-1/2} = V_{\max}$, the spectrum is continuous,
and the effective motion is infinite in one direction. For $Nm < V_{\max}$,
the spectrum is also continuous. But if we choose for the solution the
radiation condition, i.e., the solution after the barrier is taken in the
form of an outgoing wave, then this solution chooses from the continuous
spectrum separate levels which are not stationary, as in cases $\Lambda<0$
and $\Lambda = 0$.

To find these so-called quasistationary levels, we use the well-known
procedure \cite{6,7} after which we obtain the equation
$$
\sin \left( \int^{\chi_1}_0p\,d\chi + \frac\pi {4} \right)
    \exp \left( \int^{\chi_2}_{\chi_1}|p|d\chi \right)+
$$
\begin{equation}
\label{2}
    +\cos \left\{ \int^{\chi_1}_0pd\chi+\frac\pi {4} \right\}
    \exp \left\{ i\frac\pi {2}-\int^{\chi_2}_{\chi_1}
        |p| d\chi \right\}  = 0,
\end{equation}
where $|p| = \sqrt{(64\pi ^2/3)[V(\chi)-Nm]}$, $\chi_1$ and
$\chi_2$ are turning points.

Let us solve equation (\ref{2}) approximately supposing that
\begin{equation}
\label{3}
    \int^{\chi_2}_{\chi_1} |p|\, d\chi \gg 1,
    \quad\ \mbox{\rm i.e.,}\quad\
            \frac{Nm}{8\pi^2}\Lambda^{-1/2} \ll  1.
\end{equation}
Energies far less than $V_{\max}$ correspond to this case. As a
result, we obtain
$$
    Nm = [48\pi ^2(n+3/4)]^{1/2}.
$$
Writing the total energy as $Nm = E = E_0+i\Gamma$, where $E_0$ is the
energy of quasistationary levels and $1/\Gamma = \tau$ is the lifetime of
a quasistationary level with the energy $E_0$, we calculate the
penetration coefficient
$$
    D = \exp\biggl\{-2\int^{\chi_2}_{\chi_1} |p|\,d\chi\biggr\},
$$
which in this approximation has the form
$$
  D = \exp\biggl\{ -18\pi^3{\Lambda}^{-1}\biggl[ 1-\frac{(4+3\sqrt{2})}
       {\sqrt{3}}\frac{E_0\Lambda^{1/2}}{36\pi^2}\biggr] \biggr\}.
$$
In the same approximation, for the lifetime $\tau$ we have from
(\ref{2}) the following equation:
$$
    \sinh (E_0\Gamma/24\pi ) = D,
$$
or approximately $E_0\Gamma/(24\pi) \approx D$. This gives
$$
    \tau = \frac{E_0}{24\pi}\, \exp \biggl\{ 18\pi^3{\Lambda}^{-1}
    \biggl[ 1- \biggl( \frac{4+3\sqrt{2}}{\sqrt{3}} \biggr)
    \frac {E_0\Lambda^{1/2}}{36\pi^2} \biggr] \biggr\}.
$$

Let us try to compare these results with the classical ones. When
the condition (\ref{3}) is realized, there are models of two
types: oscillating (${\rm O}_1$) and monotonous (${\rm M}_2$)
\cite{8}. In the classical case, depending on the initial
conditions, only one model (${\rm O}_1$) is realized.

In the quantum approach there is a nonzero probability of a transition
from one kind of model to another, though rather small at values of $E_0$ and
$\Lambda$ characteristic of the modern epoch \cite{9}.

The same problem may be solved in the other limit when $E\approx
V_{\max} = 8\pi^2{\Lambda}^{-1/2}$. Then the barrier becomes
nearly parabolic, and the expression for the penetration
coefficient $D$ and, therefore, for the above barrier reflection
coefficient $R$ is known \cite{7}. In our case it has the form
$$
   D = \left\{1+\exp[-2\pi\Lambda^{-1/2}
        (E_0 - 8\pi^2\Lambda^{-1/2})]\right\}^{-1},
$$
or, since $|E_0-8\pi ^2{\Lambda}^{-1/2}| \ll 8\pi ^2{\lambda}^{-1/2}$,
$$
    D = \biggl\{ 1 + \exp\biggl[ -16\pi^3{\Lambda}^{-1} \biggl(
    \frac{E_0-8\pi ^2{\Lambda}^{-1/2}}{8\pi ^2{\Lambda}^{-1/2}}
    \biggr) \biggr] \biggr\}^{-1}.
$$
As it is seen from these relations, $D = 1/2$, i.e., it is not small for
$E = V_{\max} = 8\pi ^2{\Lambda}^{-1/2}$. In the classical approach, three
types of models correspond to this variant: two asymptotic models ${\rm A}_1$
and ${\rm A}_2$ and the static Einstein model. In the quantum approach the
static model is absent, but the system is described by the asymptotic models
with equal probabilities.

So, quantization of a closed isotropic model with a $\Lambda$ term leads to
new interesting results, in particular, to a finite lifetime of the system,
and questions even the fact of division of models into closed and open ones.
One should stress that these purely quantum effects appear when $\Lambda>0$,
and this is probably related to the approach interpreting the $\Lambda$ term
as vacuum energy density [9--11].

\bigskip
The authors are grateful to K.P. Staniukovich for useful discussions.



\begin{thebibliography}{99}

\bibitem{1}
    B. DeWitt, {\it Phys. Rev. } {\bf 160}, 1113 (1967).

\bibitem{2}
    V.N. Melnikov and V.A. Reshetov, {\it in:\/} Abstr. 8 All-Union
    Conference on Elementary Particles (Uzhgorod). Kiev, ITP, 1971,
    p. 117. (See also in \cite{StM}).

\bibitem{3}
    Yu.N. Barabanenkov and V.A. Pilipenko, {\it in:\/} Abstr.
    8 All-union Conference on Elementary Particles (Uzhgorod). Kiev,
    ITP, 1971, p. 117. (See also in \cite{StM}).

\bibitem{4}
    C.W. Misner, {\it Phys. Rev. } {\bf 186}, 1419 (1969).

\bibitem{5}
    Yu.N. Barabanenkov, {\it in:\/} Abstr. 3 Soviet Gravitational
    Conference, Yerevan, 1972. (See also in \cite{StM}).

\bibitem{6}
    D.I. Blokhintsev, ``Foundations of Quantum Mechanics'', Moscow,
    Vysshaya Shkola, 1963.

\bibitem{7}
    L.D. Landau and E.M. Lifshitz, ``Quantum Mechanics'', Moscow, Nauka,
    1963.

\bibitem{8}
        A.L. Zel'manov, article ``Cosmology'' in the Physical Encyclopaedic
    Dictionary.  Moscow, Sovetskaya Entsiklopediya, 1962.

\bibitem{9}
    Ya.B. Zeldovich and I.D. Novikov, ``Relativistic Astrophysics'',
    Moscow, Nauka, 1967.

\bibitem{10}
    K.A. Bronnikov, V.N.Melnikov and K.P. Stanyukovich, preprint
    ITP-68-69, Kiev, 1968. (See also in \cite{StM}).


\bibitem{11}
    K.A. Bronnikov and V.N.Melnikov, preprint ITP-69-21, Kiev,
    1969. (See also in \cite{StM}).\\
------------------------------------------------------------------
\bibitem{StM}
K.P. Staniukovich and V.N. Melnikov, ``Hydrodynamics, Fields and
Constants in the Theory of Gravitation'', Energoatomizdat, Moscow,
1983, 256 pp. (in Russian).

See English translation of first 5
sections written by V.N.Melnikov in:\\
  V.N.  Melnikov, "Fields and Constants in the Theory of Gravitation",
  CBPF MO-02/02, Rio de Janeiro, 2002, 145 pp.


\end{thebibliography}
\end{document}